\pdfoutput=1

\documentclass[12pt]{article}

\def\beq{\begin{equation}}
\def\eeq#1{\label{#1}\end{equation}}
\def\eeqn{\end{equation}}

\def\beqa{\begin{eqnarray}}
\def\eeqa#1{\label{#1}\end{eqnarray}}
\def\eeqan{\end{eqnarray}}

\let\bar=\overbar

\def\Dslash{\not{\hbox{\kern-4pt $D$}}}
\def\dslash{\not{\hbox{\kern-2pt $\del$}}}

\def\msb{{\bar{\ssstyle M \kern -1pt S}}}

\def\Title#1{\begin{center} {\Large {\bf #1} } \end{center}}

\usepackage{ifthen}
\newboolean{articletitles}
\setboolean{articletitles}{true} 

\usepackage{xspace}
\usepackage{amsmath}
\usepackage{amssymb}
\usepackage{graphicx}
\usepackage{units}
\usepackage{microtype}
\usepackage{hyperref}    
\usepackage[all]{hypcap} 
\usepackage{cite} 
\usepackage{mciteplus}
\newcommand{\CP}{\ensuremath{C\!P}\xspace}
\newcommand{\dmd}{\ensuremath{\Delta m_d}\xspace}
\newcommand{\dms}{\ensuremath{\Delta m_s}\xspace}
\newcommand{\sintwobeta}{\ensuremath{\sin 2 \beta}\xspace}

\newcommand{\cquark}{\ensuremath{c}\xspace}

\newcommand{\bquark}{\ensuremath{b}\xspace}
\newcommand{\bquarkbar}{\ensuremath{\overline \bquark}\xspace}
\newcommand{\bbbar}{\ensuremath{\bquark\bquarkbar}\xspace}

\newcommand{\pim}{\ensuremath{\pi^-}\xspace}
\newcommand{\pip}{\ensuremath{\pi^+}\xspace}
\newcommand{\mum}{\ensuremath{\mu^+}\xspace}
\newcommand{\mup}{\ensuremath{\mu^-}\xspace}
\newcommand{\kaon}{\ensuremath{K}\xspace}
\newcommand{\Kbar}{\kern 0.2em\overline{\kern -0.2em \kaon}{}\xspace}

\newcommand{\Kp}{\ensuremath{\kaon^+}\xspace}
\newcommand{\Km}{\ensuremath{\kaon^-}\xspace}
\newcommand{\Kzst}{\ensuremath{\kaon^{\ast0}}\xspace}

\newcommand{\KS}{\ensuremath{K_{\text{S}}^0}\xspace}

\newcommand{\Dzb}{\ensuremath{\kern 0.2em\overline{\kern -0.2em D}^0}\xspace}
\newcommand{\Dm}{\ensuremath{D^-}\xspace}

\newcommand{\Dsm}{\ensuremath{D_s^-}\xspace}

\newcommand{\jpsi}{\ensuremath{J\!/\!\psi}\xspace}

\newcommand{\B}{\ensuremath{B}\xspace}
\newcommand{\Bb}{\ensuremath{\kern 0.2em\overline{\kern -0.2em \B}}\xspace}
\newcommand{\Bz}{\ensuremath{\B^0}\xspace}
\newcommand{\Bzb}{\ensuremath{\Bb{}^0}\xspace}
\newcommand{\Bs}{\ensuremath{\B^0_s}\xspace}
\newcommand{\Bsb}{\ensuremath{\Bb{}^0_s}\xspace}

\newcommand{\BJpsiKS}{\ensuremath{\Bz\to\jpsi\KS}\xspace}
\newcommand{\BJpsiKst}{\ensuremath{\Bz\to\jpsi\Kzst}\xspace}
\newcommand{\BDpi}{\ensuremath{\Bz\to\Dm\pip}\xspace}
\newcommand{\BsDspi}{\ensuremath{\Bs\to\Dsm\pip}\xspace}

\newcommand{\lhcb}{LHCb\xspace}

\begin{document}

\Title{Measurements of \boldmath$\Delta m_d$, $\Delta m_s$, and $\sin 2 \beta$ \\with \lhcb}
\bigskip\bigskip


\begin{raggedright}  

{\it Julian Wishahi\index{Wishahi, J.} on behalf of LHCb collaboration\\
Technische Universit\"at Dortmund\\
Experimentelle Physik V\\
D-44221 Dortmund, GERMANY}
\bigskip\bigskip

{\it Proceedings of CKM 2012, the 7th International Workshop on the CKM Unitarity Triangle, University of Cincinnati, USA, 28 September -- 2 October 2012}
\end{raggedright}


\section{Introduction}

The LHCb experiment~\cite{external:detector} at the Large Hadron Collider (LHC)
at CERN is dedicated to the study of \bquark and \cquark flavour physics. It
exploits the large production cross-section of \bquark and \cquark hadrons in
LHC's $pp$ collisions. The measurement of time-dependent decay rates in decays
of neutral \bquark mesons, i.e.\ \Bz and the \Bs mesons, gives access to a
variety of observables that are linked to the Cabibbo-Kobayashi-Maskawa quark
mixing matrix~\cite{Cabibbo:1963yz,Kobayashi:1973fv}. The LHCb detector provides
good decay time and impact parameter resolution, and an excellent particle
identification system to efficiently reconstruct exclusive \B decay final
states.

Precision measurements of the oscillation frequencies \dmd of \Bz--\Bzb and \dms
of \Bs--\Bsb meson mixing allow to constrain the apex of the CKM triangle while
a measurement of the time-dependent \CP-asymmetry in decays of \BJpsiKS gives
access to the CKM angle $\beta$. Hence, these measurements give valuable input
to tests of the unitarity of the CKM matrix.

\section{Flavour tagging in LHCb}

Measurements of time-dependent asymmetries in decays of neutral $B$ mesons
require flavour tagging, i.e. the knowledge on the production flavour of the
reconstructed decay. As \bquark quarks are predominantly produced as \bbbar
pairs at LHCb, two classes of flavour tagging algorithms are utilised in
LHCb~\cite{internal:tagging,LHCb-CONF-2012-026}. The same-side taggers
reconstruct the charge of the hadronisation remnant of the signal $B$ meson,
i.e.\ a $K^\pm$ from the \Bs hadronisation or a $\pi^\pm$ for the \Bz/$B^\pm$. The
opposite-side taggers reconstruct the flavour of the non-signal \bquark hadron
by identifying the charge of its decay products, like leptons from semileptonic
decays or kaons from $b \to c \to s$ decays.

Inefficiencies in the tagging, e.g.\ the choice of wrong tagging particle
tracks, have to be measured in the context of \CP measurements. The tagging
efficiency $\epsilon_\text{tag}$ is the fraction of events with a tagging
decision, the mistag fraction $\omega$ represents the probability for an
incorrectly assigned tag. The overall tagging power that reflects the
statistical precision is given by $\epsilon_\text{tag}(1-2\omega)^2$.

The mistag fraction $\omega$ is calibrated on the self-tagging channel
$B^+\to\jpsi\Kp$ for the opposite-side taggers. The mixing analyses described
below are used to measure the mistag probability of the same side taggers. The
tagging power of the combined opposite side taggers is found to be around
$2.3\%$, while the same-side tagger for \Bs mesons (same-side kaon tagger) adds
about $1\%$ of tagging power and the same-side tagger for $\Bz/\B^+$ mesons
(same-side pion tagger) adds around $0.5\%$.

\section{Neutral \boldmath\B meson mixing}

Measurements of neutral \B meson mixing require a decay time dependent analysis of
the mixing behaviour. Reconstructed \B candidates are categorised by their
mixing state: they are defined as unmixed (mixed) if their production flavour
matches (does not match) the flavour at decay. This knowledge can be used to
determine the oscillation frequency $\Delta m_q$, with $q=d$ ($q=s$) in the \Bz
(\Bs) system, by using the time-dependent mixing asymmetry
\begin{align}
  \mathcal{A}_{\text{mix}}(t) = \frac{N_{\text{unmixed}}(t)-N_{\text{mixed}}(t)}
                                     {N_{\text{unmixed}}(t)+N_{\text{mixed}}(t)} 
                              = \cos\left(\Delta m_q t \right)\ ,
\end{align}
where $N_{\text{(un)mixed}}(t)$ is the number of (un)mixed candidates at \B
decay time $t$. Due to the imperfections in tagging, the accessible amplitude of
the asymmetry is reduced by $(1-2\omega)$. The amplitude is additionally damped
due to the decay time resolution. At LHCb, this dampening has to be considered
for measurements involving \Bs mixing but is negligible in \Bz mixing related
analyses.

\subsection{Measurement of \boldmath\dms in \BsDspi decays}

\begin{figure}[htb]
\begin{center}
\includegraphics[width=0.49\linewidth]{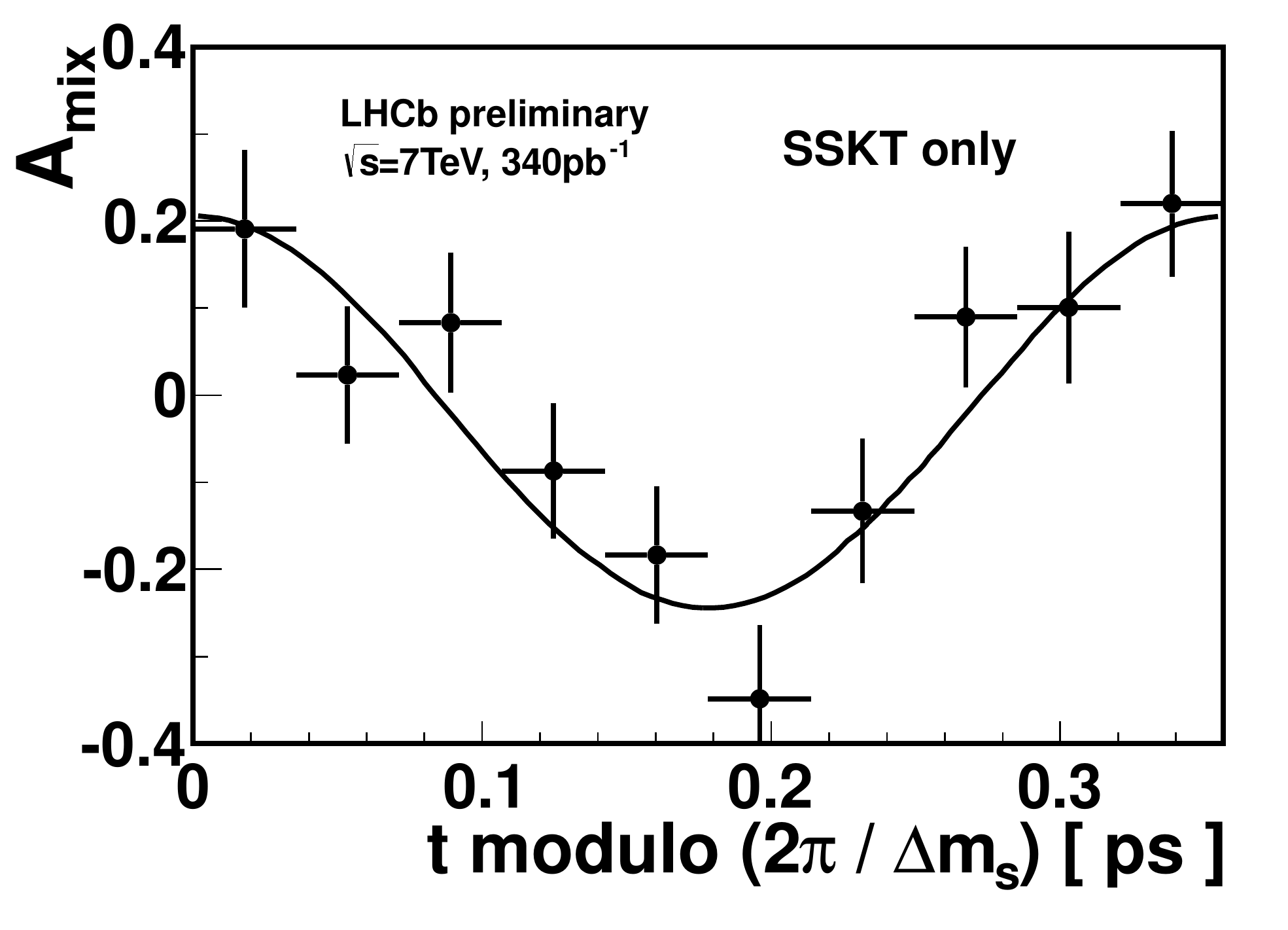}
\includegraphics[width=0.49\linewidth]{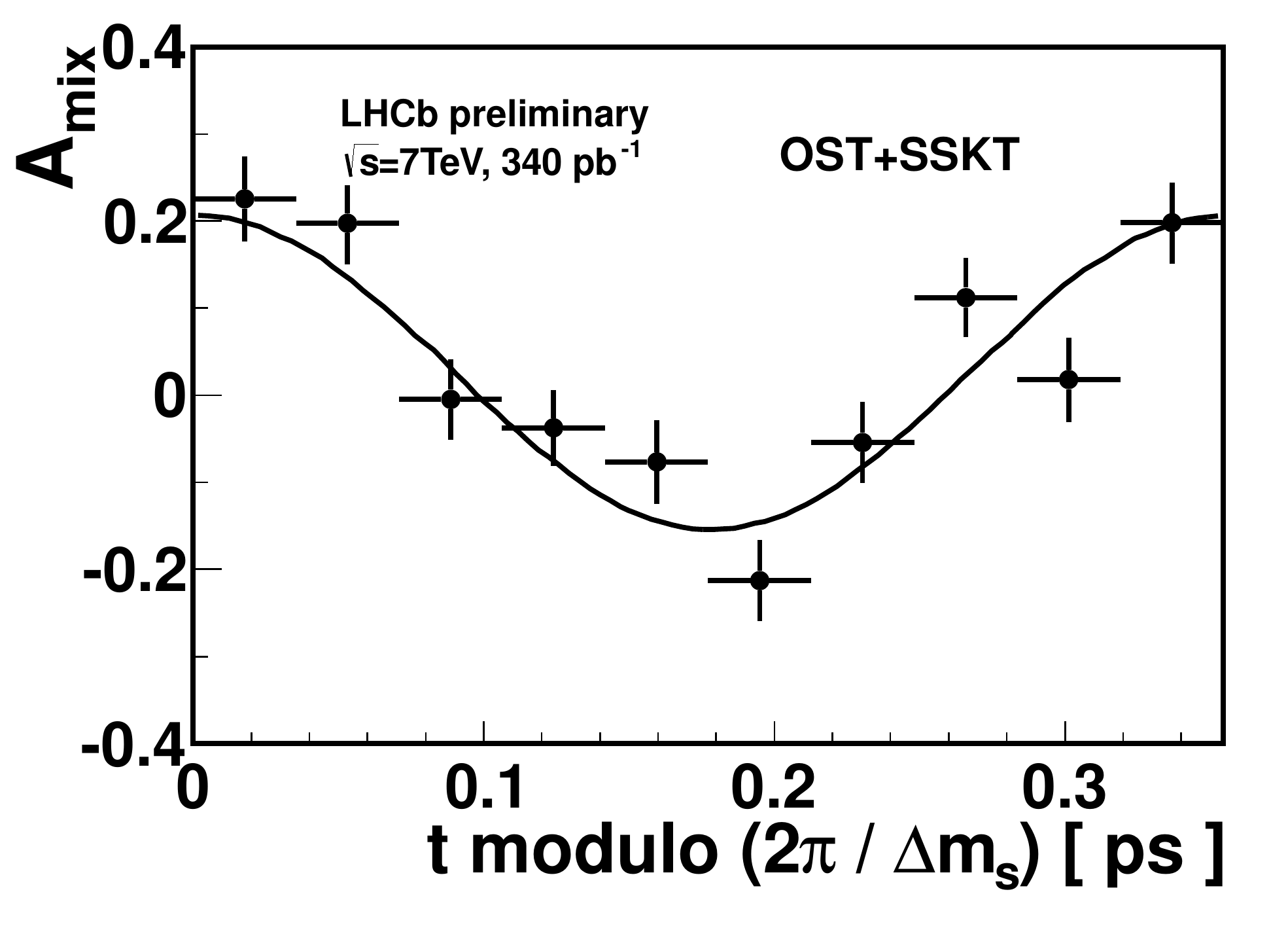}
\caption{ Mixing asymmetry for \Bs signal candidates as a function of decay time, 
          modulo $\left(\frac{2\pi}{\dms}\right)$, for the fit using only the 
          same-side tagger (left) and the combination of opposite- and same-side
          taggers (right). The fitted signal asymmetries are superimposed.}
\label{fig:dmsoscil}
\end{center}
\end{figure}

The measurement of the \Bs oscillation frequency \dms at LHCb is performed using
$\unit[340]{pb^{-1}}$ of data collected in $\sqrt{s}=\unit[7]{TeV}$ $pp$
collisions~\cite{internal:dms,conf:dms}. Around 9100 \BsDspi decays with
subsequent decays of $\Dsm\to\phi\pim$ ($\phi\to\Kp\Km$), $\Dsm\to\Kzst(892)\Km$
($\Kzst(892)\to\Kp\pim$), and the non-resonant $\Dsm\to\Kp\Km\pim$ are
reconstructed. The analysis is performed using the combination of opposite-side
taggers and the same-side kaon tagger as well as using only the same-side kaon
tagger. The resulting projection of the time-dependent mixing asymmetry is shown
in Fig.~\ref{fig:dmsoscil}. A decay time resolution of $\unit[45]{fs}$ was
measured. The analysis yields the most precise measurement of \dms with
\begin{align*}
  \dms = 17.725 \pm 0.041 \,\text{(stat.)} \pm 0.026 \, \text{(syst.)} \unit[]{ps^{-1}}\ ,
\end{align*}
where the largest systematic uncertainty is related to the uncertainty of the
length scale.

\subsection{Measurement of \boldmath\dmd in \BJpsiKst and \BDpi decays}

\begin{figure}[htb]
\begin{center}
\includegraphics[width=0.49\linewidth]{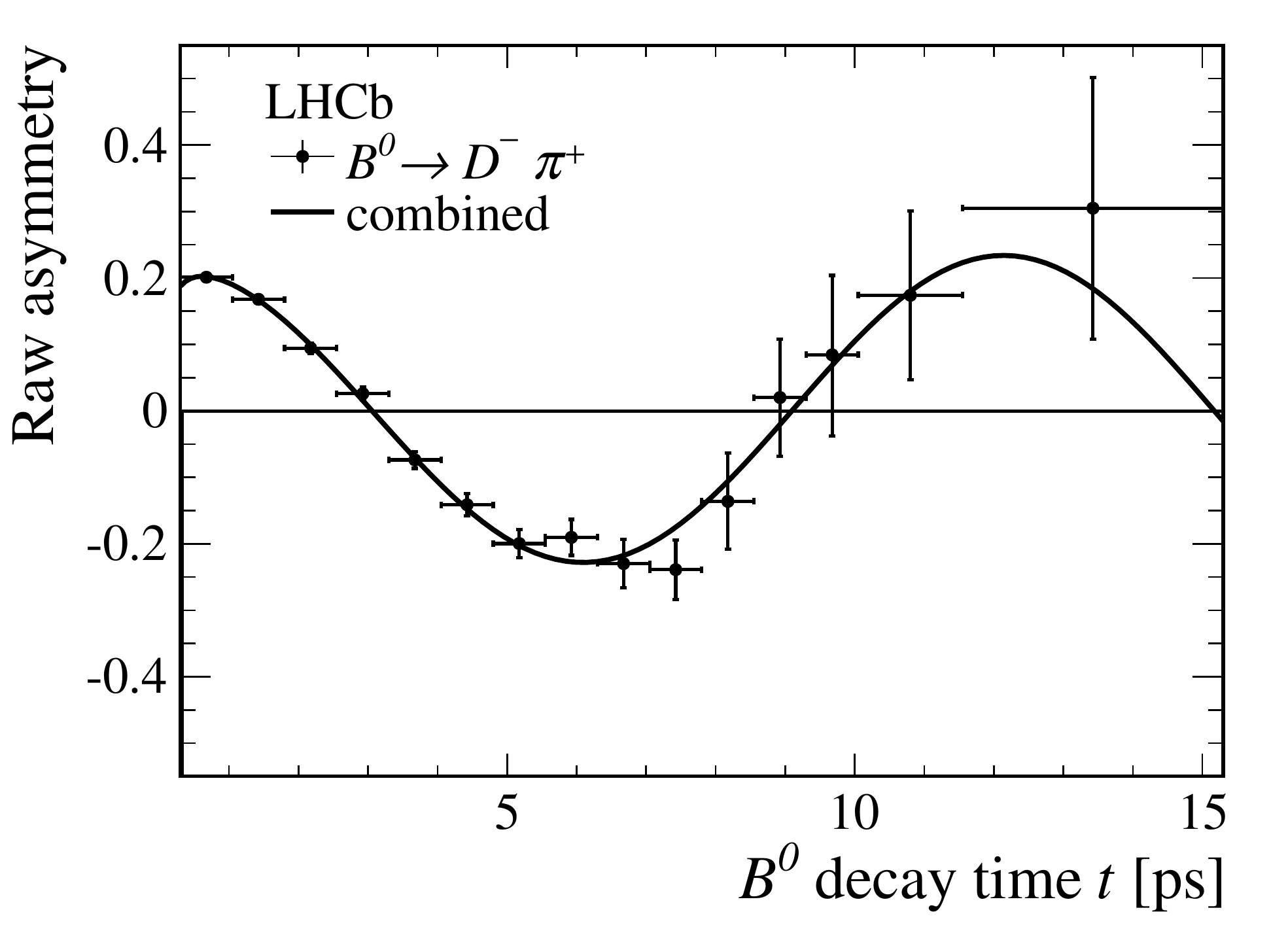}
\includegraphics[width=0.49\linewidth]{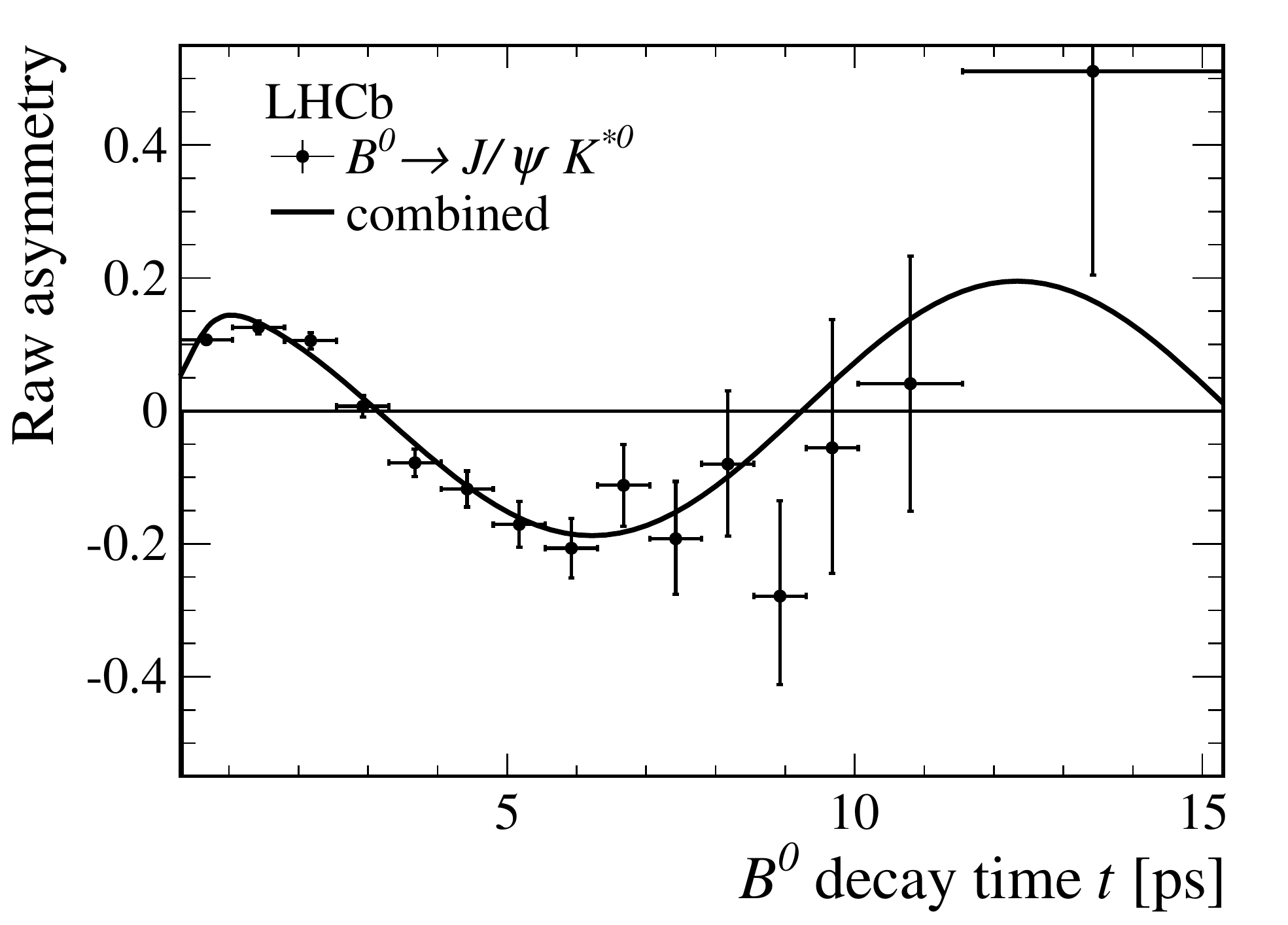}
\caption{ Raw mixing asymmetry (black points) for (left) \BDpi and (right) 
          \BJpsiKst candidates. The solid black line is the projection of the 
          mixing asymmetry of the combined PDF.}
\label{fig:dmdoscil}
\end{center}
\end{figure}

The basic analysis strategy for measuring \dmd~\cite{internal:dmd} is very
similar to the analysis strategy for the measurement of \dms, however, as the
\Bz--\Bzb oscillations are about $35$ times slower than \Bs--\Bsb oscillations,
the decay time resolution has an inferior role in this analysis. From a dataset
of $\unit[1.0]{fb^{-1}}$ collected at $\sqrt{s}=\unit[7]{TeV}$ $pp$ collisions
about $88{,}000$ decays of \BDpi ($\Dm\to\Kp\pim\pim$) and approximately
$39{,}000$ decays of $\BJpsiKst(892)$ ($\jpsi\to\mup\mum$,
$\Kzst(892)\to\Kp\pim$) are reconstructed. A combination of the opposite-side
taggers with the same-side pion tagger is used. The multi-dimensional fit to the
distributions of the reconstructed mass and the decay time yields 
\begin{alignat*}{4}
  &\dmd(\BDpi)     \, &&= 0.5178 \pm 0.0061 \, \text{(stat.)} \pm 0.0037 \, \text{(syst.)} \unit[]{ps^{-1}} \mbox{ and} \\
  &\dmd(\BJpsiKst) \, &&= 0.5096 \pm 0.0114 \, \text{(stat.)} \pm 0.0022 \, \text{(syst.)} \unit[]{ps^{-1}}.
\end{alignat*}
The mixing asymmetries in the two channels and the resulting fit projections are
shown in Fig.~\ref{fig:dmdoscil}. The largest systematic uncertainties are
related to the limited knowledge of the decay time distributions for the
background components in the fit.

Combining both channels results in the currently most precise single measurement 
of this parameter,
\begin{align*}
  \dmd = 0.5156 \pm 0.0051 \, \text{(stat.)} \pm 0.0033 \, \text{(syst.)} \unit[]{ps^{-1}}\ ,
\end{align*}
in full agreement with previous measurements~\cite{Amhis:2012bh}.

\section{Measurement of \boldmath\sintwobeta in \BJpsiKS decays}

The CKM observable \sintwobeta is one of the most precisely measured \CP
parameters. Its current world average is $\sintwobeta = 0.67 \pm 0.02 \pm
0.01$~\cite{Amhis:2012bh}, where the most precise measurements were performed
by the \B factories Belle and BaBar. 

\begin{figure}[htb]
\begin{center}
\includegraphics[width=0.7\linewidth]{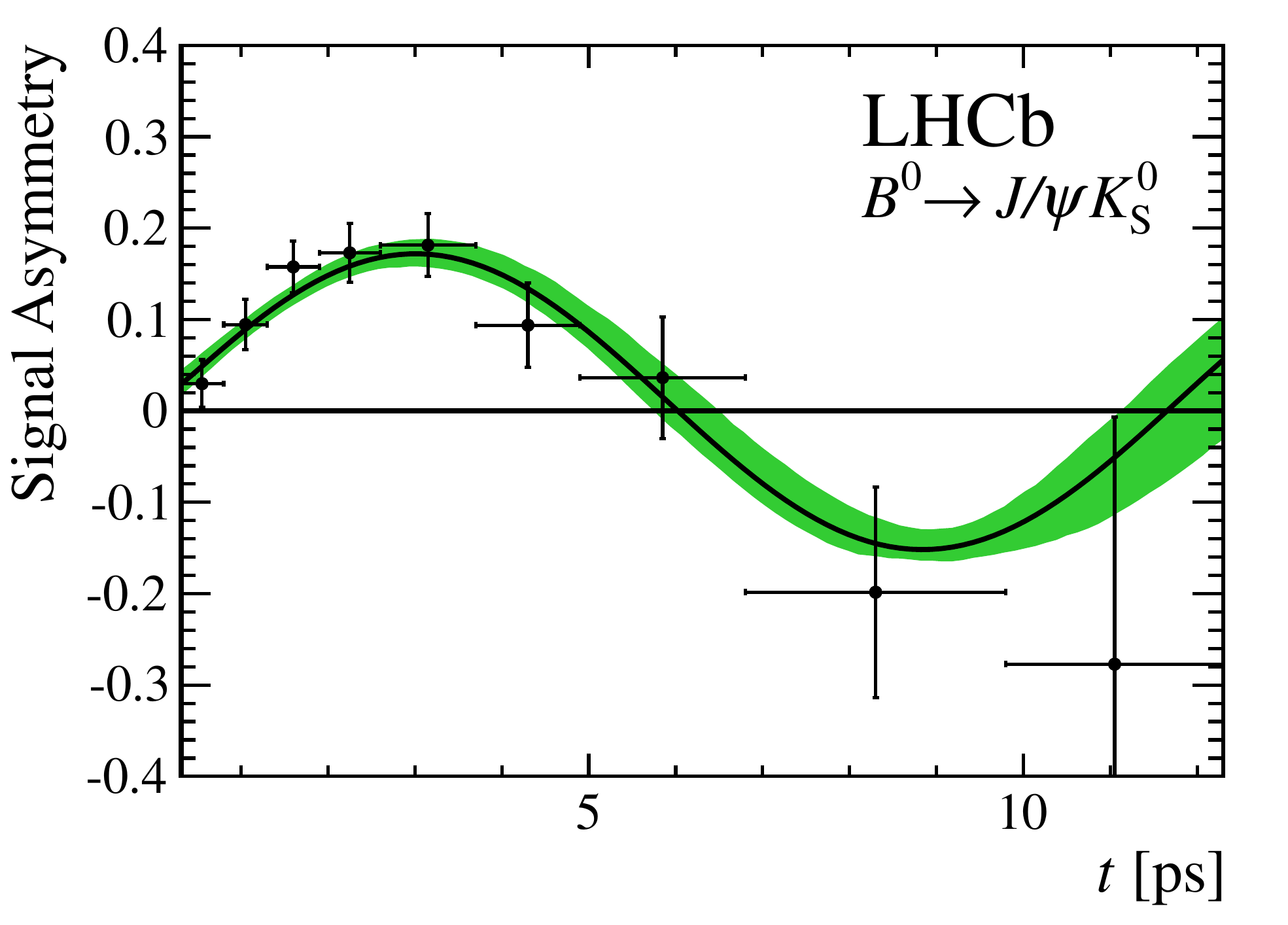}
\caption{ Time-dependent asymmetry $(N_{\Bzb}-N_{\Bz})/(N_{\Bzb}+N_{\Bz})$.
          Here, $N_{\Bz}$ ($N_{\Bzb}$) is the number of \BJpsiKS decays with
          a \Bz (\Bzb) flavour tag. The data points are obtained with the
          $_sPlot$~\cite{external:splot} technique, assigning signal weights to
          the events based on a fit to the reconstructed mass distributions. The
          solid curve is the signal projection of the PDF. The green shaded band
          corresponds to the one standard deviation statistical error.}
\label{fig:sin2b}
\end{center}
\end{figure}

For the LHCb measurement of \sintwobeta~\cite{internal:sin2b}
around $8{,}000$ \BJpsiKS ($\jpsi\to\mup\mum$, $\KS\to\pip\pim$) decays with
tagging information from the opposite side taggers are reconstructed in a data
sample of $\unit[1]{fb^{-1}}$. The parameter \sintwobeta is determined by
measuring the time-dependent \CP asymmetry
\begin{align}
  \mathcal{A}_{\jpsi\KS}(t) &= \frac{\Gamma(\Bzb(t)\to\jpsi\KS)-\Gamma(\Bz(t)\to\jpsi\KS)}
                                    {\Gamma(\Bzb(t)\to\jpsi\KS)+\Gamma(\Bz(t)\to\jpsi\KS)} \notag \\
                            &= S_{\jpsi\KS} \sin(\dmd t) - C_{\jpsi\KS} \cos(\dmd t),
\end{align}
where $S_{\jpsi\KS} = \sqrt{1-C_{\jpsi\KS}^2}\sintwobeta$. A multi-dimensional fit
of the data is performed. Fig.~\ref{fig:sin2b} shows the background corrected
asymmetry and the fit projection. 
The measurement yields
\begin{align*}
  S_{\jpsi\KS} &= 0.73 \pm 0.07 \text{\,(stat)} \pm 0.04 \text{\,(syst)} , \\
  C_{\jpsi\KS} &= 0.03 \pm 0.09 \text{\,(stat)} \pm 0.01 \text{\,(syst)} ,
\end{align*}
where the largest systematic uncertainties are related to the uncertainty of the
flavour tagging calibration and the background parameterisation. This is the
first significant measurement of \CP violation in \BJpsiKS decays at a hadron
collider. The result is in agreement with the world
averages~\cite{Amhis:2012bh}.

\section{Conclusion}

LHCb has collected $\unit[1.0]{fb^{-1}}$ of data from $pp$ collisions at a
centre-of-mass energy of $\sqrt{s}=\unit[7]{TeV}$. The time-dependent
measurements of \dms, \dmd, and \sintwobeta with this dataset demonstrate the
excellent performance of LHCb in terms of signal selection, decay time
resolution, and flavour tagging.

\end{document}